# The QE numerical simulation of PEA semiconductor photocathode

LI Xu-Dong [1,2]  GU Qiang[1]  ZHANG Meng[1]  ZHAO Ming-Hua[1]

[1] Shanghai Institute of Applied Physics, Chinese Academy of Sciences, Shanghai 201800, China

[2] Graduate University of the Chinese Academy of Sciences, Beijing 100049, China

**Abstract:**

Several kinds of models have already been proposed for explaining the photoemission process. The exact photoemission theory of semiconductor photocathode was not well established after decades of research. In this paper an integral equation of quantum efficiency (QE) is constructed to describe the photoemission of positive electron affinity (PEA) semiconductor photocathode based on three-step photoemission model. The influences of forbidden gap, electron affinity, photon energy, incident angle, degree of polarization, refractive index, extinction coefficient, initial/final electron energy, relaxation time and external electric field on the QE of PEA semiconductor photocathode are taken into account. In addition, a computer code is also programmed to calculate the QE of $K_2CsSb$ photocathode theoretically at 532nm wavelength, the result is in line with the experimental value by and large. What are the reasons caused to the distinction between the experimental measuring and theoretical QE are discussed.

# 1 Introduction

Generally speaking, metallic and semiconductor photocathodes [1] are routinely used in high brightness, low transverse emittance and high average current photoinjector. Metallic photocathodes have prompt response time, insensitive to contamination, easy to manufacture, accept high electric field, small electron beam emittance and dark current, but the poor quantum efficiency (QE) with ultraviolet (UV) radiation. In contrary, semiconductor photocathodes have different characteristics with the highest QE at longer wavelength, however, sensitive to contamination, slow response time.

The bialkali photocathodes can be commonly used in photocathode electron gun since they have higher QE, lower emittance and dark current than other photocathodes. $K_2CsSb$ [2] is the most typical bialkali photocathode, which is operated under the visible light irradiate, that means more available power and stable drive laser system.

The researchers make their utmost effort to discover and prepare the superior photocathode, Meanwhile, whose devote themselves to develop a successful photoemission theory for various photocathode. So far, the most successful photoemission model is three-step photoemission model, Although the model does not include any the surface-specific effects, assume transport to the photocathode-vacuum surface and escape from the surface are independent process, which has been successfully used to interpret the photoelectron emitting process of elemental metal, metallic compound, alkali antimonide, alkali telluride, alkali halide, III-V photocathode [3-6], etc.. In the following, the approximate QE equation of reflection mode PEA semiconductor photocathode based on the previous theoretical and experimental results is obtained and verified.

## 2 Derivate the QE equation of PEA semiconductor photocathode

In semiconductor energy band model, $h$ is the plank constant, $v$ is the laser frequency, $E_{VBM}$ is the valence band maximum, $E_{CBM}$ is the conduction band minimum, $E_{VAC}$ is the vacuum level, $E_F$ is the Fermi level, $E'_{VAC}$ is the vacuum level after the schottky effect reduction. $\Delta E_a$ is the external electric field-induced electron affinity decrease, $E_g$ is the forbidden band gap which cannot be occupied by electrons, and $E_a$ is the electron affinity which is an interval between the conduction band minimum and the vacuum level, as shown in Fig. 1.

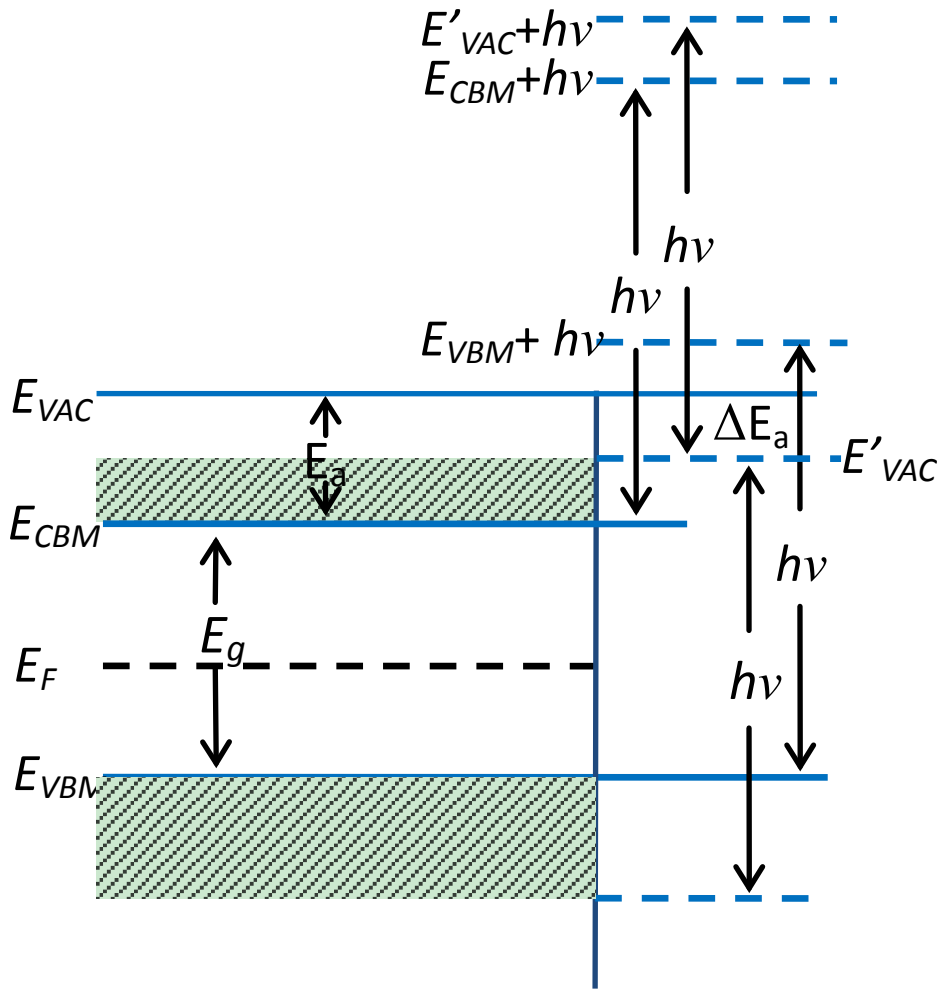

Fig. 1. The simplified energy band diagram of PEA semiconductor photocathode.

It is well known that semiconductor photocathode in the external accelerating electric field is illuminated with photons energy greater than the sum of band gap and electron affinity, these bound electrons in the valence band and conduction band will absorb photon energy, and then emit from the interior to vacuum as photoelectrons. The spicer's three-step photoemission model is used to explain the process, there are three independent processes in the simple model: photon absorption and photoelectrons excitation in the photocathode, photoelectrons migration to the photocathode-vacuum interface and photoelectrons with sufficient energy escape from the photocathode surface to produce electron beam. Each step of the model has an associated probability. QE is just the product of the probabilities in each step of photoemission, which is simply given by

$$QE(hv) = \int_{E_{bottom}}^{E_{top}} P(E, hv, T) F_\lambda(s, E, v, \theta) D(E) dE, \tag{1}$$

where $E$ is the electron energy; $E_{bottom}$, $E_{top}$ are the bottom and top electron energy of the valence/conduction band in semiconductor, respectively; $P(E,hv,T)$ is the probability that electron from the initial state of energy $E$ to the final state of energy $E+hv$, $T$ is photocathode temperature, the unit is K; $F_\lambda(s,E,v,\theta)$ is the probability that photoelectron arrives at the photocathode surface undergoing several scattering, $s$ is the depth, $\theta$ is the angle of the photoelectron velocity relative to the surface normal; $D(E)$ is the probability that photoelectron jumps into vacuum.

*Step 1: Absorb photons and excite electrons*

The first step in the three-step photoemission model is these photoelectrons are determined by the number of photons inside the photocathode that can be absorbed and the number of electrons that can be excited. We hypothesize the photoelectrons are from the valence band and conduction band in semiconductor photocathode respectively, the probability $P(E,h\nu,T)$ is solely the function of $\rho_{EDOS}(E)$, $f_{FD}(E,T)$, $\rho_{EDOS}(E+h\nu)$, and $f_{FD}(E+h\nu,T)$, which can be written as

$$P(E,h\nu,T) = [[1-R(n,k,i)] \times \frac{\rho_{EDOS}(E+h\nu)\big(1-f_{FD}(E+h\nu,T)\big)\rho_{EDOS}(E)f_{FD}(E,T)}{\int_{E_{VBM}}^{E_{VBM}+h\nu} dE'\,\rho_{EDOS}(E'+h\nu)\big(1-f_{FD}(E'+h\nu,T)\big)\rho_{EDOS}(E')f_{FD}(E',T)} \quad (2)$$

$$\frac{+\rho_{EDOS}(E+h\nu)\big(1-f_{FD}(E+h\nu,T)\big)\rho_{EDOS}(E)f_{FD}(E,T)}{+\int_{E_{CBM}}^{E_{CBM}+h\nu} dE'\,\rho_{EDOS}(E'+h\nu)\big(1-f_{FD}(E'+h\nu,T)\big)\rho_{EDOS}(E')f_{FD}(E',T)},$$

the denominator represents a normalization based on the assumption that the 100% absorption of all photons, so that the integral of $P(E,h\nu,T)$ over all possible values of $E+h\nu$ will yield unity.

As everyone knows the electron distribution of semiconductor photocathode material is Fermi-Dirac distribution

$$f_{FD}(E,T) = \frac{1}{1+exp\left(\frac{E-E_F}{k_B T}\right)}, \quad (3)$$

here $k_B$ is the Boltzmann constant, $E_F$ is the Fermi level, the approximate formula is

$$E_F = \frac{E_{VBM}+E_{CBM}}{2}. \quad (4)$$

The absorption of photons is affected by the reflectivity $R(n,k,i)$ and the absorbing coefficient, $i$ is the incident angle. In general, the reflectivity is a function of wavelength, incident angle, polarization state of incident light, refractive index, extinction coefficient and photocathode thickness. In the process of theoretical deduction, the reflectivity of photocathode is related to the complex refractive index by the generalized Fresnel reflection coefficients, $n$ is the real (i.e., refraction index) and $k$ the imaginary part (i.e., extinction coefficient) of $\eta$ (i.e., complex refraction index). $n$ and $k$ are also relative to photon energy, with $n$ and $k$ in hand, the reflectivity as a function of polarization state are obtained from Ref.[7]

$$R = \frac{1}{2}\left[R_p(1+p) + R_s(1-p)\right], \quad (5)$$

here $R_s$ is the component of reflection perpendicular to the plane of incidence, $R_p$ is the parallel component, $R$ is the reflectivity of incident light. The degree of polarization for the incident light is given by $p = (I_p - I_s)/(I_p + I_s)$, $I_p$ and $I_s$ are the parallel and perpendicular component of light intensity respectively.

$\lambda_{opt}(\nu)$ is the penetration depth (i.e., skin depth), which is also as a function of extinction coefficient and refraction index, $\lambda_{opt}(\nu)$ can be usually simplified as the inverse of the absorption coefficient $\alpha$,

$$\lambda_{opt}(\nu) = 1/\alpha = \lambda/4\pi nk, \quad (6)$$

where $\lambda$ is the incident photon wavelength.

*Step 2: migrate to surface*

The second step involves diffusion of these photoelectrons to the photocathode surface. The assumption of the exponential model about the photoelectron transport is the average effect of plenty of nearly elastic and inelastic collisions when the photocathode thickness is much greater than photoelectron scattering length. The probability of the total scattering depends on photon frequency, electron energy and $\theta$. The fraction of photoelectrons per unit distance at a depth s below the surface, which survive scattering is

$$F_{\lambda}(s,E,\nu,\theta)=\frac{\int_0^{\infty} exp\left(-\frac{s}{\lambda_{opt}(\nu)}-\frac{s}{l(E)cos\theta}\right)ds}{\int_0^{\infty} exp\left(-\frac{s}{\lambda_{opt}(\nu)}\right)ds}=\frac{cos\theta}{cos\theta+\sqrt{\frac{m}{2E}}\frac{\lambda_{opt}(\nu)}{\tau(E)}}, \quad (7)$$

where $m$ is the effective electron mass, which is related to the band gap via $m/m_0=E_g/R_{\infty}$, $R_{\infty}$ is the Rydberg energy 13.606 eV, $m_0$ is the free electron mass. $l(E)$ is the scattering length,$l(E)=\upsilon(E)\tau(E)=\sqrt{2E/m}\,\tau(E$, $\upsilon(E)$ is the electron velocity, $\tau(E)$ is relaxation time, The photoelectrons may lose energy and change direction of movement in the process of photoelectrons diffusion inside the material towards the photocathode material-vacuum interface by collision with other electrons (electron-electron scattering), lattices (electron-phonon scattering) and impurities (electron-impurity scattering). The general diffusion model has been presented by Kane, which includes the scattering length from electron-electron, electron-phonon and other scattering interactions. According to the above discussion and $\tau(E)$ follows the Matthiessen's Rule with the relationship of various scattering [8]

$$\tau_{total}^{-1}=\sum_j \tau_j^{-1}. \quad (8)$$

Due to the small number of free electron of the conduction band in semiconductor, the photoelectron moving towards surface-vacuum interface, electron-electron scattering is negligible. Electron-impurity scattering effect on photoelectron transmission can be minimized by reducing impurity concentration, optimizing crystal size and orientation. The main scattering is dominated by electron-phonon scattering (the energy loss through electron-phonon scattering is rather small, which will mainly change the moving direction of photoelectron). It is obvious that the electron-phonon scattering contains polar optical phonon scattering and acoustic phonon scattering. The previous researchers [8] have found that the contribution of acoustic phonon scattering for the relaxation time is small compared to polar optical phonon scattering. Therefore the impact of polar optical phonon scattering on relaxation time is only considered. The equation of relaxation time originate from optical phonon scattering is analogous to $Cs_3Sb$ photocathode (The format is shown as in Ref.[8]). The differences are the high frequency dielectric constant ($\varepsilon_{\infty}$) is given approximately by the Drude theory expression, $\varepsilon_{\infty}=n^2$ and the static dielectric constant $\varepsilon_0\approx n^2+k^2$. So

$$\Delta\varepsilon=1/n^2-1/(n^2+k^2). \quad (9)$$

*Step 3: Escape from the surface*

The third step is these photoelectrons do undergo all kinds of scattering events, only those photoelectrons with energy component directed into surface barrier are greater than surface barrier height can overcome surface barrier, and then enter into vacuum. In order to simulate it, the perpendicular component of the photoelectron momentum must satisfy [3]

$$\frac{\hbar^2 k_{\perp}^2}{2m}>E_{VBM}+E_g+E_a-\Delta E_a, \quad (10)$$

where $\Delta E_a$ is the field-induced electron affinity decrease of semiconductor photocathode by Schottky effect. Under neglecting the band bending of internal electric field conditions, in view of the properties of semiconductor photocathode and the equation of metallic Schottky effect, the reduction of electron affinity due to Schottky effect is described by the following equation in semiconductor photocathode

$$\Delta E_a = \sqrt{\frac{\varepsilon - 1}{\varepsilon + 1}\beta q^3 E_D}, \tag{11}$$

where $\varepsilon$ is the dielectric constant of semiconductor photocathode, $\varepsilon \approx \varepsilon_0$ for low frequency electric field; $q$ is the elemental electron charge; $E_D$ is the external electric field strength, $E_D = V/d$, $V$ is the applied voltage and $d$ is the distance between the cathode and anode; $\beta$ is the dimensionless effective field enhancement factor of photoemission.

In order to describe the photoelectrons escape from interior to vacuum. The maximum emission angle, $\theta_{max}(E)$, of the photoelectrons with enough energy can enter into vacuum as follows

$$\theta_{max}(E) = cos^{-1}\left(\sqrt{\frac{E_{VBM} + E_g + E_a - \Delta E_a}{E + h\nu}}\right), \tag{12}$$

Eq. (12) is suitable for describing the photoelectron emission of valence band and conduction band. When $\theta_{max}(E)$=0 to $\pi$, these photoelectrons may move to any direction of cathode internal. For $\theta_{max}(E)>0$, The probability that photoelectrons enter into vacuum is given by

$$D(E) = \frac{\frac{1}{4\pi}\int_0^{\theta_{max}(E)} sin\,\theta d\theta \int_0^{2\pi} d\varphi}{\frac{1}{4\pi}\int_0^{\pi} sin\,\theta d\theta \int_0^{2\pi} d\varphi} = \frac{1}{2}\int_0^{\theta_{max}(E)} sin\,\theta d\theta. \tag{13}$$

From Eq. (1), Eq. (2), Eq. (7) and Eq. (13), the QE equation is

$$\mathrm{QE} = [1 - \mathrm{R}(n,k,i)] \times \frac{\int_{E_{VBM}+E_g+E_a-\Delta E_a-h\nu}^{E_{VBM}} dE\,A(E,h\nu,T)\int_0^{\theta_{max}} \sin\theta d\theta \int_0^{\infty} \exp\left(-\frac{s}{\lambda_{opt}(\nu)} - \frac{s}{l(E)\cos\theta}\right) ds \int_0^{2\pi} d\varphi}{\int_{E_{VBM}+E_g+E_a-\Delta E_a-h\nu}^{E_{VBM}} dE\,A(E,h\nu,T)\int_0^{\pi} \sin\theta d\theta \int_0^{\infty} \exp\left(-\frac{s}{\lambda_{opt}(\nu)}\right) ds \int_0^{2\pi} d\varphi}$$
$$+ \frac{\int_{E_{VBM}+E_g}^{E_{VBM}+E_g+E_a-\Delta E_a} dE\,A(E,h\nu,T)\int_0^{\theta_{max}} \sin\theta d\theta \int_0^{\infty} \exp\left(-\frac{s}{\lambda_{opt}(\nu)} - \frac{s}{l(E)\cos\theta}\right) ds \int_0^{2\pi} d\varphi}{+\int_{E_{VBM}+E_g}^{E_{VBM}+E_g+E_a-\Delta E_a} dE\,A(E,h\nu,T)\int_0^{\pi} \sin\theta d\theta \int_0^{\infty} \exp\left(-\frac{s}{\lambda_{opt}(\nu)}\right) ds \int_0^{2\pi} d\varphi}, \tag{14}$$

here

$$A(E,h\nu,T) = \rho_{EDOS}(E + h\nu)\big(1 - f_{FD}(E + h\nu,T)\big)\rho_{EDOS}(E)f_{FD}(E,T). \tag{15}$$

When the cathode temperature is low, $k_B T \ll E_F$, the Fermi-Dirac distribution is well represented by the Heaviside-step function [4].

$$\int_{bottom}^{E_{top}} dE\,\big(1 - f_{FD}(E + h\nu,T)\big)f_{FD}(E,T) = \int_{bottom}^{E_{top}} dE. \tag{16}$$

Substitute Eq.(16) into Eq. (14) and calculate, the QE equation is simplified as

$$QE = [1 - R(n,k,i)] \times \frac{\int_{E_{VBM}+E_g+E_a-\Delta E_a-h\nu}^{E_{VBM}} B(\nu,\theta,E)C(E)dE + \int_{E_{VBM}+E_g}^{E_{VBM}+E_g+E_a-\Delta E_a} B(\nu,\theta,E)C(E)dE}{2\int_{E_{VBM}+E_g-h\nu}^{E_{VBM}} C(E)dE + 2\int_{E_{VBM}+E_g}^{E_{VBM}+E_g+E_a-\Delta E_a} C(E)dE}, \tag{17}$$

here

$$B(\nu,\theta,E)=\int_{\cos\theta_{max}\ (E)}^{1}\frac{\cos\theta}{\cos\theta+\sqrt{\frac{m}{2E}}\frac{\lambda_{opt}\,(\nu)}{\tau(E)}}d(\cos\theta)=1-\sqrt{\frac{m}{2E}}\frac{\lambda_{opt}\,(\nu)}{\tau(E)}\ \times$$

$$l\,n\left[1+\sqrt{\frac{m}{2E}}\frac{\lambda_{opt}\,(\nu)}{\tau(E)}\right]-\cos\theta_{max}\ +\sqrt{\frac{m}{2E}}\frac{\lambda_{opt}\,(\nu)}{\tau(E)}\,l\,n\left[\cos\theta_{max}\ +\sqrt{\frac{m}{2E}}\frac{\lambda_{opt}\,(\nu)}{\tau(E)}\right], \tag{18}$$

$$C(E)=E^{1/2}(E+h\nu)^{1/2}.$$

## 3 The QE numerical Simulation of $K_2CsSb$ photocathode based on the QE equation

The experimental measuring results of *n* and *k* by D.Motta and S.Hallensleben [9,10] are shown in Fig. 2. *n* and *k* are related to incident wavelength and photocathode thickness. Table I also shows $n$=3.3 and $k$=0.8, when thin $K_2CsSb$ photocathode is irradiated with 532nm light.

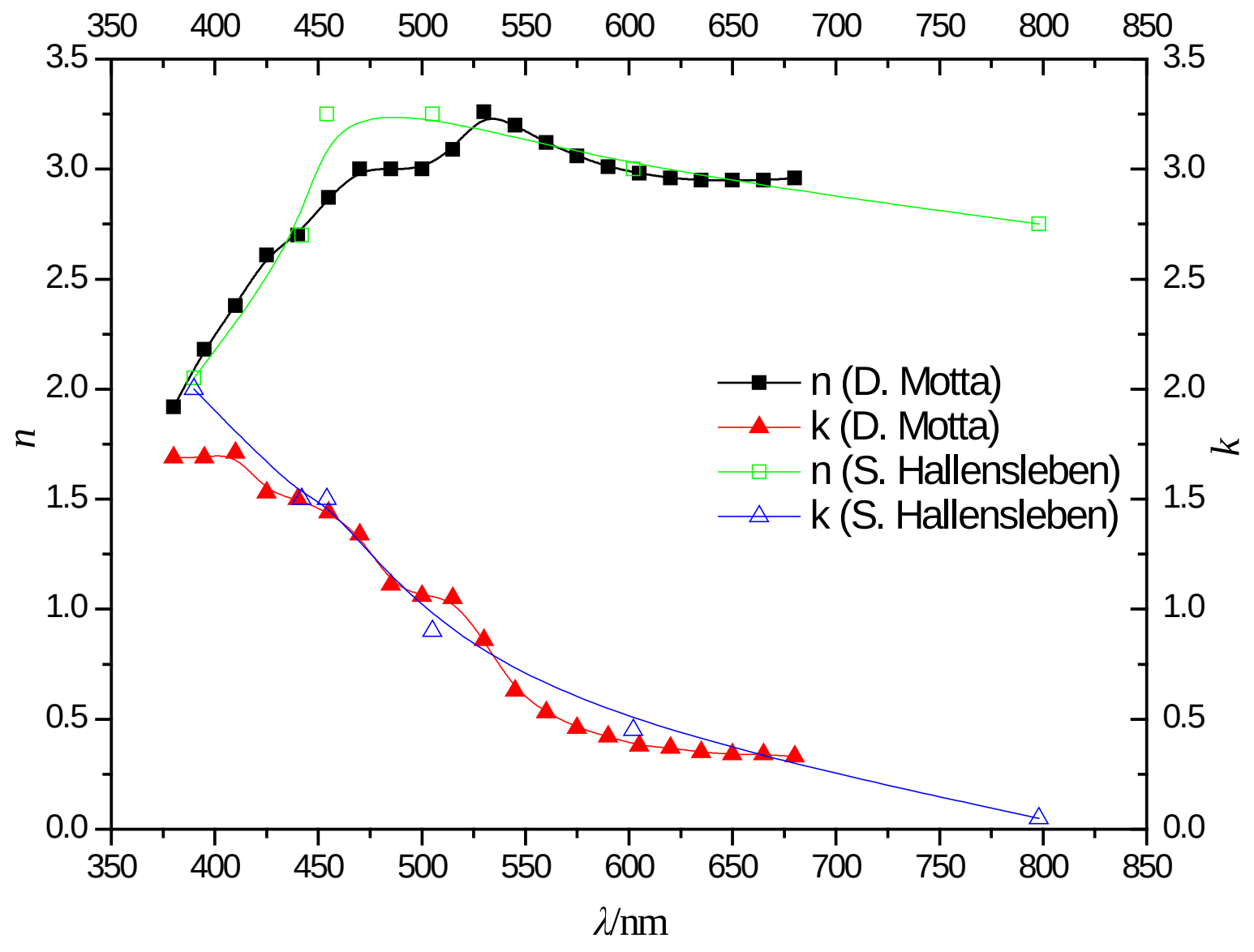

Fig. 2. The refractive index and extinction coefficient versus wavelength of $K_2CsSb$ photocathode.

The sound velocity $v_s$ is inferred from the elastic constant $c_{11}$ and the mass density $\rho$ by the relationship [8] $v_s = \sqrt{c_{11}/\rho}$. Choosing $\rho$ to be the ratio of the mass of atoms in a unit cell with the unit cell volume $L^3$ of $K_2CsSb$, $L$ is lattice constant. The $K_2CsSb$ crystal structure is the $DO_3$ cubic structure ($O_h^5$ space group [11]).The unit cell contains four $K_2CsSb$ units. That is

$$\rho = \frac{4(2m_K + m_{Cs} + m_{Sb})}{L^3} = 3.4643\, g/cm^3. \quad (19)$$

For $m_{Cs}$=2.21×10$^{-22}$ g ($m_{Cs}$ is the mass of each cesium atom), $m_{Sb}$=2.02×10$^{-22}$ g, $m_K$=6.49×10$^{-23}$ g and $c_{11}$ is a “generic” semiconductor value of 12×10$^{11}$ N/m$^2$. The phonon energy$\hbar\omega_q$ is crudely related to the sound velocity in the relation suggested by Ridley of $\hbar\omega_q \approx 4\pi\hbar v_s/L \approx 0.5654$ eV, the value is slightly larger than $Cs_3Sb$. $\omega_q \approx 4\pi v_s/L$=8.5899×10$^{13}$Hz, $\omega_q$ is phonon frequency. These parameters in simulation are given in Table I.

Table I. Parameters used in simulation for $K_2CsSb$ photocathode.

| Parameter | Symbol | Value | Unit |
|---|---|---|---|
| Band gap | $E_g$ | 1.2 | eV |
| Electron affinity | $E_a$ | 0.7 | eV |
| Valence band maximum | $E_{VBM}$ | 1.27 | eV |
| Optical wavelength | $\lambda$ | 532 | nm |
| Incident angle | $i$ | $\pi/12$ | … |
| Refractive index | $n$ | 3.3 | … |
| Extinction coefficient | $k$ | 0.8 | … |
| Degree of polarization | $p$ | -99/101 | |
| Temperature | $T$ | 300 | K |
| Electric field | $E_D$ | $2\times10^4$ | V/m |
| Field enhancement factor | $\beta$ | 1 | … |
| lattice constant | $L$ | 8.61 | Å |
| Sound velocity | $v_s$ | 5885.5 | m/s |
| Plank constant | $h$ | $4.136\times10^{-15}$ | J·s |
| Light velocity | $c$ | $3\times10^8$ | m/s |
| Rydberg constant | $R_\infty$ | 13.606 | eV |
| Electron charge | $q$ | $1.60\times10^{-19}$ | C |
| Boltzmann constant | $k_B$ | $8.62\times10^{-5}$ | eV/K |
| Permittivity of vacuum | $\varepsilon_0$ | $8.85\times10^{-12}$ | F/m |

# 4 Results and Discussions

*QE* =4.69%, when 532nm light was used to irradiate the $K_2CsSb$ photocathode. The theoretical simulation is in agreement with the experimental value in some research institutes. For example, photocathode with QE of 3% at 532nm wavelength light irradiated in BNL [12] and Cornell University [13]. When $K_2CsSb$ photocathode is evaporated on Mo and stainless steel (SS) substrates, the maximum QE measured is typically 6% at 532 nm in LBNL [14]. Nevertheless, the agreement between theoretical and experimental QE is not so excellent? These reasons will be listed as follows

Firstly, as a matter of fact, the drive light is not absolutely monochromatic. Secondly, the stoichiometry of alkali antimonide photocathode is varying with the photocathode depth, so that the composition of K-Cs-Sb photocathode is not really $K_2CsSb$. The QE of alkali-antimonide compound is measured in experiment. Thirdly, the absorption coefficient for $K_2CsSb$ photocathode has been calculated with the incident photon energy changing by L. Kalarasse [15], which is different from the theoretical equation, $\alpha = 4\pi nk/\lambda$. Actually, the absorption coefficient is not calculated from $n$, $k$ and $\lambda$, which only can be measured precisely by experimental method. Fourthly, other scattering effects on relaxation time have been omitted, only the polar photon scattering has been considered. Fifthly, the Fermi-Dirac distribution is not precisely represented by the Heaviside-step function at the non-absolute zero. Sixthly, the uncertainly of the Fermi level of $K_2CsSb$ gives rise to the deviation. Finally, the measuring forbidden band gap and electron affinity of $K_2CsSb$ photocathode are different by many researchers [16] as shown in Table II.

Table II. The forbidden band gap and electron affinity of $K_2CsSb$ photocathode in different institutes.

| | $E_g$(eV) | $E_a$(eV) |
|---|---|---|
| NRL | 1.1 | 0.65 |
| BARC | 1.2 | 0.7 |
| BNL | 1.0 | 1.1 |

The dependence of QE on the angle of incident light, polarization state of incident light, the applied bias voltage and photocathode temperature with certain photon energy has also been simulated is shown in

Fig. 3. It is found that the QE of $K_2CsSb$ photocathode changed with the incident angle of light quickly. The influences of photocathode temperature, degree of polarization of incident light and electric field on QE can be neglected under certain conditions.

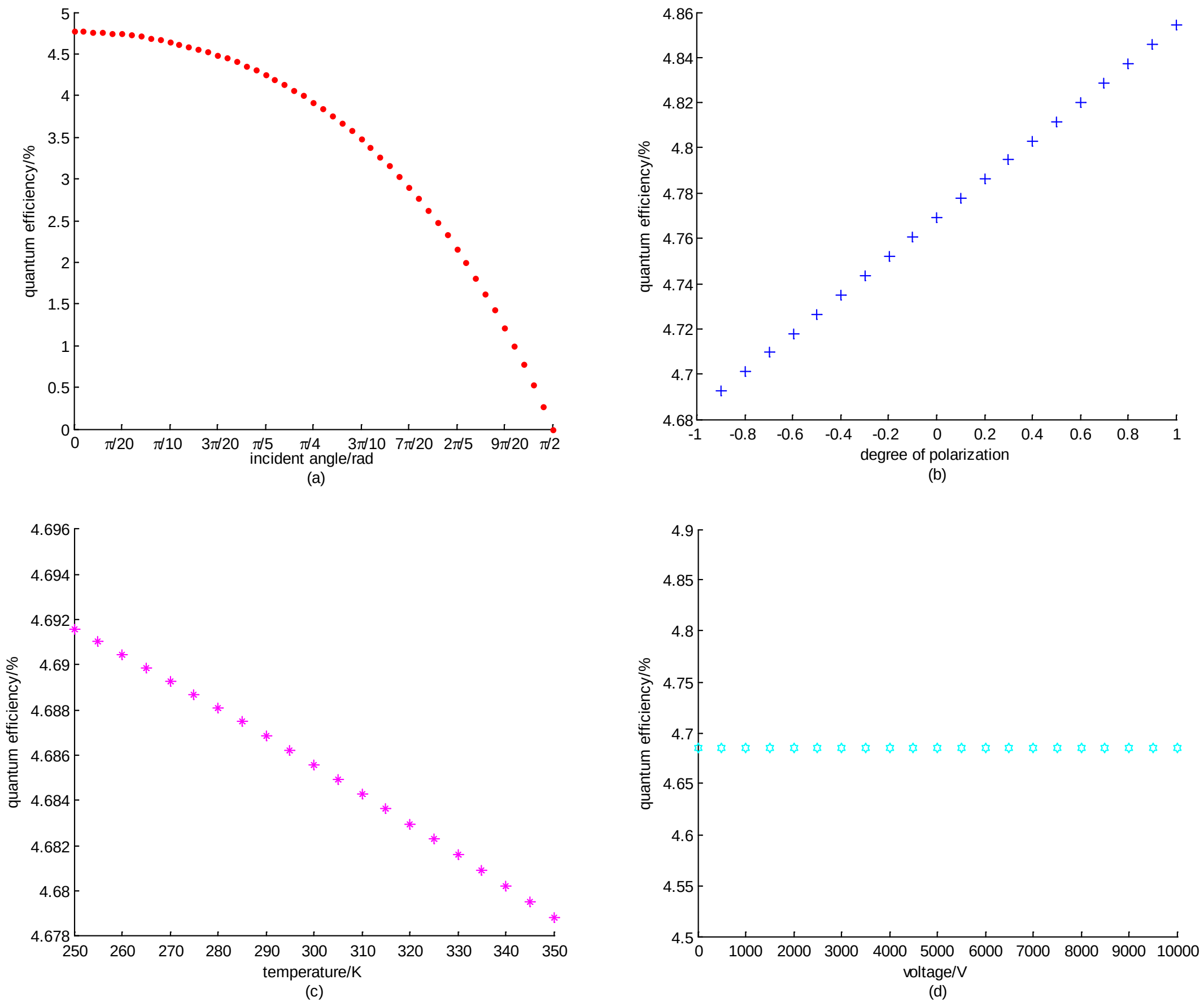

Fig. 3. QE as a function of incident angle of light (a), degree of polarization of light (b), photocathode temperature (c) and the external accelerating electric field (d) at 532nm wavelength for $K_2CsSb$ photocathode respectively.

## 5 Conclusions

One appropriate model was developed and verified to be capable of explaining photoemission process of PEA semiconductor photocathode in simulation. The probability of each step of three-step photoemission model has been considered, photon absorption, photoelectron diffusion and photoelectron escape are embodied by the QE equation. The simulative QE is 4.69% from computer code, simulation also indicated that the different effects of incident angle, polarization state, photocathode temperature and external electric field on QE. These research results could be useful not only to the study of pre-existing photocathode, but also to the discovery of the new high efficient photocathode. In the near future, we shall update the relative parameters in accordance with our experimental condition.

## 6 Acknowledgements

The authors would like to thank Erdong Wang of BNL for his beneficial discussions about the theory of photoemission.